\documentclass[12pt,paper]{JHEP}
\usepackage{epsf}
\newcommand{\ee}{\end{equation}}
\newcommand{\be}{\begin{equation}}
\newcommand{\ea}{\end{eqnarray}}
\newcommand{\ba}{\begin{eqnarray}}

\newcommand{\rpiV}{\langle r^2\rangle^\pi_V}

\title{The Vector and Scalar Form Factors of the Pion to Two Loops}

\author{J. Bijnens\\Dept. of Theor. Phys. 2, Univ. Lund,\\
S\"olvegatan 14A, S--22362 Lund, 
Sweden}
\author{G. Colangelo\\INFN--Laboratori Nazionali di Frascati,\\ P. O. Box
  13, I--00044 Frascati, Italy}
\author{P. Talavera,\\Dept. de F{\'\i}sica i Enginyeria Nuclear,
UPC,\\E-$08034$ Barcelona, Spain}
\abstract{We calculate the vector and scalar form factors of the pion to
  two loops in Chiral Perturbation Theory. We estimate the
  unknown ${\cal O}(p^6)$ constants using resonance exchange. We make a
  careful comparison to the available data and determine two ${\cal
    O}(p^4)$ constants rather precisely, and two ${\cal O}(p^6)$ constants
  less precisely. We also use Chiral Perturbation Theory to two
  loops to extract in a model--independent manner the charge radius of the
  pion from the available data, and obtain $\rpiV=0.437\pm0.016$~fm$^2$.}

\preprint{LU TP 98-12\\May 1998}
\begin{document}
\section{Introduction}
Chiral Perturbation Theory (CHPT) \cite{CHPTreviews}
is the modern way of exploiting Chiral
Symmetry constraints on strongly interacting processes. The mesonic
two--flavour sector was treated in an extensive work to
next--to--leading order \cite{GL1}. The physical processes and amplitudes
treated in that reference were $F_\pi$, $M_\pi^2$, the scalar and vector
form factors of the pion, the pion radiative decay ($\pi\to \ell\nu\gamma$)
and $\pi\pi$ scattering. 
Due to recent and planned experimental improvements, more accurate
theoretical analyses are needed. 
The next--to--next--to--leading order of $F_\pi$, $M_\pi^2$ were calculated 
in \cite{Buergi,BCEGS2}, the pion radiative decay in \cite{BT} and
$\pi\pi$ scattering in \cite{BCEGS2,BCEGS1}. In addition, the pion
polarizabilities and the production of pion pairs in gamma--gamma collisions
are also known to this order. For the latter, the neutral process  
was studied in \cite{BGS} and the
charged case in \cite{Buergi}. In this paper we present the full 
${\cal O}(p^6)$ calculation of the scalar and vector pion form factors,
thereby completing the calculation to the next order of the processes
considered in \cite{GL1}.

In addition to the above calculations there are those where the amplitude
is calculated using dispersive methods. This method allows for the full
calculation up to the subtraction constants. These have then to be determined
by comparison with experiment. The disadvantage of this approach is that
one cannot do a simple comparison to
existing models of low--energy
constants appearing in CHPT, since the ``chiral logarithm'' parts of
the subtraction constants cannot be fully
determined\footnote{Sometimes
the requirement that the limit $M_\pi^2\to 0$ gives a finite result allows
to determine these terms, see \cite{GM} for an example.}. As a consequence,
we can neither vary the quark mass to compare with e.g. lattice calculations.
The dispersive calculation for the form factors was done in \cite{GM},
(see \cite{Knecht} for the analogous $\pi\pi$ calculation)
and an analytical expression for the dispersive integrals in
the vector form factor was found in \cite{tau}.
The one--loop formula and the latter partial two--loop results for the vector 
form factor
have been used in the inverse amplitude method to fit
over a larger kinematic range \cite{Hannah}. A different resummation scheme
using constraints from Vector Meson Dominance (VMD), 
$1/N_c$ and unitarity, works as well
in extending the range of validity \cite{Guerrero}.
For a recent review of the form factors in general see 
Ref. \cite{Drechsel}.

We have performed the calculation in two different ways. We have used
the master formula approach as described in \cite{BCEGS2} and we have also
directly computed all the relevant Feynman diagrams.
Several nontrivial checks on the calculation of both form factors were done.
The first is the absence of nonlocal divergences, i.e. the dependence
on quark masses and external kinematical variables of the divergent
parts must be analytic. Another powerful check results from the value of the
form factors at zero momentum transfer.
For the vector form factor this must be one because of
gauge invariance and, for the scalar form factor, it is related to
the derivative of the pion mass w.r.t. the quark mass. The latter
relation follows from the Feynman--Hellman theorem.

For both form factors we have abundant experimental information to compare
with. In the vector case this information consists of direct measurements
of the form factor, both in the timelike and spacelike region. In the
scalar case, since there is no microscopic scalar probe available, it is
impossible to directly measure the form factor. On the other hand, it can
be shown that the experimental information on the scalar, $I=0$
$\pi \pi$ phase shifts, inserted in a dispersive representation for the
form factor, allows one to completely reconstruct the energy dependence of
the form factor, modulo a multiplicative overall factor.
Therefore in both cases we can make a detailed comparison of the CHPT
two--loop expressions to the available experimental information.
The fact that the form factors have a simple kinematical
structure, makes even the two--loop representation rather simple and
easy to manipulate. For this reason the comparison to the experimental
information is particularly instructive.

The contributions at order $p^6$ can be split into two
different pieces: a dispersive contribution and a polynomial part. 
The numerical contribution of the dispersive part has been already analyzed
in Ref. \cite{GM}.
Inside the polynomial part we have again two types of terms: chiral
logarithms and new ${\cal O}(p^6)$ low energy constants (LEC). The splitting
between these two types of terms is arbitrary and depends on the
renormalization scale $\mu$: on the other hand we have learned from the
experience at order $p^4$ that the $\rho$ mass scale   
is a sensible choice for understanding the physical origin of
these two types of terms. As we will show in our analysis, the same choice
seems to be still sensible at order $p^6$, since we will be able to
understand at least the order of magnitude and sign of the new LEC's with
an estimate based on the resonance saturation hypothesis. 
We will also confirm at order $p^6$ the Vector Meson Dominance hypothesis,
showing the different importance of the resonance contribution in the
scalar and in the vector channel.
Concerning the vector form factor, we stress that the direct comparison of 
CHPT to the data allows a reliable, model--independent extraction of the
value of the charge radius.

The paper is organized as follows. In Sect. \ref{definitions}
we define the form factors and the notation used. In Sect. \ref{calculation}
we describe the calculation and the checks performed, and give
the analytical results for the form--factors and the associated
radii. Sect. \ref{resonance} describes various estimates of the ${\cal
  O}(p^6)$  constants appearing in the calculation. In  
Sect. \ref{numerics} we describe the other numerical input used and
fit accurately to the available data for both form factors. Here we also
describe the variation of the fits with several different assumptions
and provide numerical results for all quantities, including $F_\pi/F$,
$M_\pi^2/M^2$ and an improved estimate of the low energy
hadronic vacuum polarization contribution to the muon anomalous
magnetic moment. 
In this section we also discuss the Omn\`es representation.
We finally recapitulate our main conclusions in
Sect. \ref{conclusions}.
  
\section{Definitions and Notation}
\label{definitions}

\subsection{Form factors}
The scalar and vector form factors are defined respectively by
\ba
\label{defFVFS}
\langle \pi^i(p_2)| \bar u u + \bar d d | \pi^j(p_1)\rangle &=&
 \delta^{ij} F_S(s_\pi)  \; ,
\nonumber\\
\langle \pi^i(p_2)|\frac{1}{2}\left( \bar u\gamma_\mu u - \bar d\gamma_\mu d
\right) | \pi^j(p_1)\rangle &=&
 i \varepsilon^{i3j} (p_{1\mu}+p_{2\mu}) F_V(s_\pi)\; ,
\ea
where $s_\pi = (p_2-p_1)^2$. The scalar form factor is defined through the
isospin--zero scalar source. The isospin--one scalar form factor can be
defined analogously but it only starts at ${\cal O}(p^4)$. The vector form
factor is also an isovector, and what we calculate here is its $I_z=0$
component. Similar definitions exist for the other isospin components. 
In what follows we take the isospin limit i.e. $m_u=m_d=\hat{m}$, so that
the other isospin--one components of the vector form 
factors are identical to the one we analyze here. 

\subsection{Chiral Perturbation Theory}
Quantum chromodynamics with two flavours in the chiral limit has
an $SU(2)_L\times SU(2)_R\equiv O(4)$ symmetry which is spontaneously
broken down to $SU(2)_V\equiv O(3)$. At low energies, the three resulting
Goldstone Bosons are the only relevant degrees of freedom and their
interactions are strongly constrained by the underlying symmetry. The three
Goldstone Bosons can be identified with the pions and the way of extracting
the consequences of the chiral symmetry is Chiral Perturbation Theory.
We use the non--linear sigma model or $O(4)$ parametrization with the
external field formalism of Ref. \cite{GL1}.

To lowest order in the low energy expansion, ${\cal O}(p^2)$, 
processes are described by the tree--level diagrams of the lagrangian
\begin{equation}
{\cal L}_2 = \frac{F^2}{2} \nabla_\mu U^{\dagger} \nabla_{\mu} U +
 F^2 (\chi^T U),
\end{equation}
with the covariant derivative defined by
\ba
\nabla_{\mu} U^0 &=& \partial_\mu U^0 + a^i_\mu (x) U^i,\nonumber \\
\nabla_\mu U^i &=& \partial_\mu U^i + \epsilon ^{i k l} v^k_\mu(x) U^l
- a^i_\mu (x) U^0,
\ea
where $U(x)$ is an $O(4)$ four--component vector
\be
U^T= \left(\sqrt{1-\frac{\pi^i\pi^i}{F^2}},\frac{\pi^1}{F},\frac{\pi^2}{F},
\frac{\pi^3}{F}\right).
\ee
Here $v^k(x)$ and $a^i(x)$ are the external vector and axial--vector
sources respectively
and $\chi = 2 B (s^0,p^i)$ contains the isospin--zero part of the
scalar source and the isospin--one part of the pseudoscalar source.

To ${\cal O}(p^4)$ the amplitudes are given by one--loop graphs with vertices
from ${\cal L}_2$ and tree--level graphs containing vertices from ${\cal
  L}_2$ and one vertex from the ${\cal O}(p^4)$ Lagrangian given by
\begin{eqnarray}
{\cal L}_4 &=& l_1 (\nabla^\mu U^\dagger \nabla_\mu U )^2
+ l_2 (\nabla^\mu U^\dagger \nabla^\nu U)(\nabla_\mu U^\dagger
\nabla_\nu U) \nonumber \\
& &+ l_3 (\chi^\dagger U)^2 + l_4 (\nabla^\mu \chi^\dagger
\nabla_\mu U) + l_5 (U^\dagger F^{\mu \nu } F_{\mu \nu} U) \nonumber \\
& &+ l_6 (\nabla^\mu U^\dagger F_{\mu \nu} \nabla^\nu U) + l_7 (
\tilde{\chi}^\dagger U)^2 + h_1 \chi^\dagger \chi + h_2 F_{\mu \nu}
F^{\mu \nu} \nonumber \\
& & + h_3 \tilde\chi^\dagger \tilde\chi,
\end{eqnarray}
where the strength tensor $F_{\mu \nu}$ is defined by
\begin{equation}
(\nabla_\mu \nabla_\nu - \nabla_\nu \nabla_\mu) U = F_{\mu \nu} U\,,
\end{equation}
and $\tilde\chi=2 B (p^0,s^i)$.

The ${\cal O}(p^6)$ contributions contain pure 
two--loop diagrams with vertices from
${\cal L}_2$, one--loop diagrams with vertices from ${\cal  L}_2$ and
one vertex from ${\cal L}_4$, and tree level diagrams with vertices
from ${\cal L}_2$ and either two vertices from ${\cal L}_4$ or
one vertex from the ${\cal O}(p^6)$ Lagrangian. The latter Lagrangian,
${\cal L}_6$, can be found in \cite{Fearing}.

\section{The calculation}
\label{calculation}

The calculation has been made using two different methods: in one case, we
have calculated directly all the relevant Feynman diagrams with full
generality. 
As an alternative method we use the master equation approach
\cite{BCEGS2}, which corresponds
to recognizing that a large part of the graphs comes together such that
they are one--loop graphs with one of the vertices given either by the
one--loop scalar or vector form factor or by the $\pi\pi$ scattering
amplitude with the pion legs off--shell. Some two--loop diagrams with
non--overlapping loops have then to be evaluated separately in order to
obtain the correct normalization. 
The integrals have been evaluated
using the methods of \cite{BuGS}, see also \cite{Buergi,BCEGS2,BT}.
The subtraction procedure we used is a version of the modified
minimal subtraction ($\overline{\mbox{MS}}$) as described in \cite{BCEGS2}.

Both methods have been used independently and yielded the same result.
They also satisfy the requirements of gauge invariance, i.e. $F_V(0)=1$,
and of the Feynman--Hellman theorem, i.e.
$F_S(0) = \partial M_\pi^2 /\partial \hat{m} $. Both constraints
couple quite a few diagrams in each process 
providing thus a good check on the calculations. 
In addition, at two--loop order there are nonlocal divergences diagram
by diagram, which also cancel in the sum. The ``double
chiral logs'' also satisfy the constraints imposed by renormalization, 
meaning that the terms of the type 
$\left(\log({M_\pi^2}/{\mu^2})\right)^2$
can always be cast inside the $k_i$ quantities defined
in Sect. \ref{FpiMpi}-- see \cite{Gilberto} for further 
explanation and references.
 

\subsection{$F_\pi$ and $M_\pi^2$}
\label{FpiMpi}

For the sake of completeness we write down here
the pion decay constant and the pion mass. This also serves
to introduce the notation we use for the other quantities of interest. The
difference with the expressions used in \cite{Buergi,BCEGS2} is that we
have subtracted the infinities using $\overline{\mbox{MS}}$ scheme
and rewritten the ${\cal O}(p^4)$ part in terms of the 
physical mass and decay constant of the pion ($M_\pi^2$~ and~$ F_\pi$ 
respectively).
\ba
\label{FpiF}
\frac{F_\pi}{F} &=& 1 + x_2 (l_4^r-L)
+x_2^2\Bigg[
\frac{1}{N} \left(-\frac{1}{2}l_1^r -l_2^r+\frac{29}{12}L\right)
  -\frac{13}{192}\frac{1}{N^2}
\nonumber\\&& + \frac{7}{4}k_1 + 
k_2 - 2 l_3^r l_4^r + 2(l_4^r)^2 - \frac{5}{4}k_4 + r_F^r\Bigg]+{\cal O}(x_2^3)
\, ,
\ea
and
\ba
\label{MpiM}
\frac{M_\pi^2}{M^2} &=& 1 + x_2 (2 l_3^r+\frac{1}{2}L)
+x_2^2\Bigg[\frac{1}{N}\left( l_1^r + 2 l_2^r  -\frac{13}{3}L\right)
  + \frac{163}{96}\frac{1}{N^2}
\nonumber\\&&
 -\frac{7}{2}k_1 - 2 k_2 - 4(l_3^r)^2 + 4l_3^r l_4^r - \frac{9}{4} k_3 
  + \frac{1}{4} k_4 +r_M^r\Bigg]+{\cal O}(x_2^3)\,.
\ea
The constants $r_F^r$ and $r_M^r$ denote the contributions from
the ${\cal O}(p^6)$ lagrangian after modified minimal subtraction.
 
We have defined the following quantities
\ba
\label{defvarious}
N&=&16\pi^2\,,\nonumber\\
x_2&=&\frac{M_\pi^2}{F_\pi^2}\,,\nonumber\\
L &=&\frac{1}{N}\log\frac{M_\pi^2}{\mu^2}\,,\nonumber\\
k_i&=& (4 l_i^r-\gamma_i L)L\,,\nonumber\\
M^2 &=& 2 B \hat{m}\,,
\ea
$M^2$ being the lowest order pion mass and $F$ the pion decay constant
in the chiral limit.
The $l_i^r$ are the finite part of the coupling constants $l_i$ in
${\cal L}_4$ after the $\overline{\mbox{MS}}$ subtraction, and their
values depend on the renormalization scale $\mu$ as
$\mu^2(d l_i^r/d\mu^2)= -(\gamma_i/2N)$. The $\gamma_i$ were
calculated in \cite{GL1} and are given by
\be
\gamma_1=\frac{1}{3};~\gamma_2=\frac{2}{3};~\gamma_3=-\frac{1}{2};~
\gamma_4=2;~\gamma_5=-\frac{1}{6};~\gamma_6=-\frac{1}{3};~\gamma_7=0\,.
\ee
Later we will also follow common use, and discuss the  ${\cal O}(p^4)$
LEC's in terms of their scale--invariant combinations, the $\bar{l}_i$'s,
defined as
\be
l_i^r =\frac{\gamma_i}{2N}(\bar{l}_i+N L) \; \; .
\ee

\subsection{Scalar Form Factor}

We start with the expression of the scalar form factor to two
loops evaluated with the methods described above. 
\ba
\label{FS}
F_S(s)&=& F_S(0)
\left\{1+x_2\left(\frac{1}{2}(2s-1) \bar{J}(s)+
s (l_4^r-L-\frac{1}{N}) \right)+x_2^2 \left({\it P_S^{(2)}}+{\it U_S^{(2)}}
\right) \right\}\nonumber\\&&+{\cal O}(x_2^3)\,.
\ea
As has been already mentioned, the two--loop contribution has been
split into two parts:
the polynomial piece of the amplitude reads
\ba
\lefteqn{ P_S^{(2)}(s) =  s^2 \Bigg[
 - \frac {11}{12} k_1 - \frac{7}{12} k_2 - \frac{1}{4} k_4} \nonumber\\
&&+\frac{1}{N} \left(\frac{7}{1728}  - \frac{32}{9} l_1^r - 
 \frac {19}{9} l_2^r -  l_4^r + 
\frac {85}{36}L \right) +\frac {1817}{1296} \frac {1}{N^2} +r_{S3}^r\Bigg]
\nonumber\\ & & + s \Bigg[\frac {31}{6} k_1 +\frac {17}{6} k_2- k_4 
+\frac {1}{N}\left( -\frac {11}{864} + \frac {110}{9} l_1^r
+ \frac {40}{9}l_2^r - 2 l_4^r + \frac {11}{9}L \right) \nonumber\\
&& - \frac{20}{81}\frac{1}{N^2}+ 2(l_4^r)^2 -4 l_3^r l_4^r+{\it r_{S2}^r}\Bigg]
\,,
\ea
where
\[
s = s_\pi/M_\pi^2 \, \, ;
\]
while the dispersive piece can be cast in the following form
\ba
\lefteqn{ U_S^{(2)}(s) =  \bar{J}(s) \Bigg[ 
\frac{1}{3} l_1^r\,(11 s^2 - 40 s + 44) + \frac {1}{3} l_2^r\,(7 s^2
 - 20 s + 28)}\nonumber \\
&&  + 5 l_3^r + l_4^r \left( s^2 + \frac{3}{2} s - 1 \right)  + 
\frac{1}{18}L\, \left( - 43 s^2
 + 53 s - \frac {119}{2} \right) \nonumber \\
&& + \frac {1}{N} \left(\frac{29}{12}s^2 -\frac {61}{9}s + 
\frac {391}{36} \right)\Bigg]+
\frac {3}{4} K_1(s) + K_2(s)\,\left( \frac {43}{36} s^2 - 
\frac {4}{3}s +\frac {1}{4} \right) \nonumber \\
&&  + K_3(s)\, \left(\frac{1}{3}s - \frac{25}{18} \right)\,.
\ea

The integral functions $\bar J, K_1, K_2,K_3$ and $K_4$
are defined in \cite{BCEGS2} and we reproduce them here for the sake of
completeness 
\be
\left(\begin{array}{l}\bar{J}\\
K_1\\
K_2\\
K_3\\
\end{array}\right)
=
\left(\begin{array}{cccc}
0&0&z&-4N\\
0&z&0&0\\
0&z^2&0&8\\
Nzs^{-1}&0&\pi^2(Ns)^{-1}&\pi^2\\
 \end{array} \right)
 \left(\begin{array}{c}
{h}^3\\
{h}^2\\
{h}\\
\displaystyle{-(2N^2)^{-1}}
\end{array}
\right)
\ee
and
\be
K_4=\frac{1}{sz}\left(\frac{1}{2}K_1+\frac{1}{3}K_3+
\frac{1}{N}\bar{J}
+\frac{(\pi^2-6)s}{12N^2}\right)\,,
\ee
where
\be
{h}(s)=\frac{1}{N\sqrt{z}}\ln
\frac{\sqrt{z}-1}{\sqrt{z}+1} \quad ,\qquad z=1-\frac{4}{s} \; .
\ee
The functions $s^{-1}\bar{J}$ and $s^{-1}K_i$ are analytic in the complex
$s$--plane (cut along
the positive real axis for $s \geq 4$), and they vanish
as $|s|$ tends to infinity. Their real and
imaginary parts are continuous
at $s=4$.

We use the form factor at zero momentum transfer given by
\ba
\label{FS0}
\lefteqn{ F_S(0) =2B\Bigg\{1 + x_2\,\left( 4 l_3^r + L+\frac{1}{2N} \right) }
\nonumber\\
& & + x_2^2\Bigg[
\frac {1\,}{N} \left( -11 l_1^r -2 l_2^r -3 l_3^r+l_4^r -\frac{39}{4}L \right)
 +  \frac{97}{96}\frac{1}{N^2}\nonumber \\
 && - 8(l_3^r)^2 + 8 l_3^r l_4^r -\frac{21}{2}
k_1 - 6 k_2 -\frac{21}{4} k_3 +\frac{1}{2} k_4 + r_{S1}^r\Bigg]\Bigg\}
+{\cal O}(x_2^3)\,,
\ea
as a check of the previous result, Eq. (\ref{FS}).
One can see that this last expression can also be derived 
using Eq. (\ref{MpiM}) and the Feynman--Hellman theorem. It agrees exactly
and leads to the relation between the ${\cal O}(p^6)$ constants
$r_{S1}^r = 3 r_M^r$.

We can now expand the form factor for $s\ll 1$  ($s_\pi\ll 4M_\pi^2$),
obtaining the expression
\be
\label{FSexpansion}
F_S(s)= F_S(0)\left( 1 + \frac{1}{6}\langle r^2\rangle^\pi_S s + c_S^\pi s^2
+{\cal O}(s^3)\right)\,.
\ee
This serves as the definition of the pion scalar radius
$\langle r^2\rangle^\pi_S$ and of the coefficient $c^\pi_S$.
Expanding the integral functions in Eq. (\ref{FS}) we obtain
\ba
\label{rpi2s}
\lefteqn{ \langle r^2 \rangle^\pi_S = 
x_2\left(-\frac{13}{2N}+6 l_4^r - 6L\right)}\nonumber\\
&&+ x_2^2\Bigg[\frac {1}{N} 
 \left(  -\frac{23}{192} + 88 l_1^r
 + 36 l_2^r + 5 l_3^r - 13 l_4^r +
\frac{145}{36}L  \right)  + \frac{869}{108}\frac{1}{N^2}
\nonumber \\
&& - 24 l_3^r l_4^r + 12 (l_4^r)^2 + 31 k_1 + 17 k_2 -6 k_4 + 6 r_{S2}^r
\Bigg]+{\cal O}(x_2^3)\,, 
\ea
and an analogous formula for $c_S^\pi$
\ba
\label{cpis}
c_S^\pi&=& x_2 \frac{1}{N} \frac{19}{120}+ x_2^2\Bigg[
\frac{1}{N} \left(\frac{5}{1152} - \frac{83}{15}l_1^r
 -\frac{46}{15}l_2^r +\frac{ 1}{12}l_3^r -\frac{23}{30}l_4^r
          + \frac{6041}{2160}L\right)\nonumber\\&&
+\frac{1655}{1296}\frac{1}{N^2}
  -\frac{11}{12}k_1 - \frac{7}{12}k_2 - \frac{1}{4}k_4 +r_{S3}^r\Bigg]
+{\cal O}(x_2^3)\,.
\ea

\subsection{Vector Form Factor}

As in the previous subsection we start with the general
expression for the form factor. In the vector case it is given by
\be
\label{FV}
F_V=1+x_2 \Bigg[ \frac{1}{6}(s-4)\bar{J}(s)+s
\left(-l_6^r-\frac{1}{6}L-\frac{1}{18N}\right) 
\Bigg]+x_2^2\bigg(P_V^{(2)}+U_V^{(2)}\bigg)+{\cal O}(x_2^3)\,.
\ee
One should notice that at zero momentum transfer gauge invariance 
constrains this
form factor to be $F_V(0)=1$.

Once more we find it instructive to split the different contributions to the
form factor.
The polynomial part of the amplitude is
\ba
\lefteqn{
P_V^{(2)}=
s^2\Bigg[\frac{1}{12}k_1-\frac{1}{24}k_2+\frac{1}{24}k_6}\nonumber\\&&
+\frac{1}{9N}\left(l_1^r-\frac{1}{2}l_2^r+\frac{1}{2}l_6^r
-\frac{1}{12}L-\frac{1}{384}
-\frac{47}{192N}\right)+r_{V2}^r\Bigg]\nonumber\\&&
+s\Bigg[-\frac{1}{2}k_1+\frac{1}{4}k_2-\frac{1}{12}k_4+\frac{1}{2}k_6
-l_4^r\left(2l_6^r+\frac{1}{9N}\right)
\nonumber\\&&
+\frac{23}{36}\frac{L}{N}+\frac{5}{576N}
+\frac{37}{864N^2}+r_{V1}^r
\Bigg]\,,
\ea
and the dispersive part of the form factor is
\ba
U_V^{(2)}&=&
\bar{J}(s)
\Bigg[\frac{1}{3} l_1^r (-s^2+4s) +\frac{1}{6} l_2^r(s^2-4s)
+\frac{1}{3}l_4^r(s-4)+\frac{1}{6}l_6^r(-s^2+4s)\nonumber\\&&
-\frac{1}{36}L(s^2+8s-48)+\frac{1}{N}\left(\frac{7}{108}s^2
-\frac{97}{108}s+\frac{3}{4}\right)\Bigg]+\frac{1}{9}K_1(s)\nonumber\\&&
+\frac{1}{9} K_2(s)\left(\frac{1}{8}s^2-s+4\right)
+\frac{1}{6}K_3(s)\left(s-\frac{1}{3}\right)-\frac{5}{3}K_4(s)\,.
\label{UV}
\ea
We can now expand the form factor for $s\ll 1$ ($s_\pi\ll 4M_\pi^2$)
and obtain the expression
\be
\label{FVexpansion}
F_V=1+\frac{1}{6}\langle r^2 \rangle^\pi_V s+c^\pi_V s^2+{\cal O}(s^3)\,,
\ee
where the pion charge radius is then given by
\ba
\label{rpi2vchpt}
\langle r^2 \rangle^\pi_V = &&
x_2\left(-6l_6^r-L-\frac{1}{N}\right)
+x_2^2\Bigg[-3k_1+\frac{3}{2}k_2-\frac{1}{2}k_4+3k_6
-12l_4^r l_6^r\nonumber\\&&
+\frac{1}{N}\left(-2l_4^r+\frac{31}{6}L+\frac{13}{192}
-\frac{181}{48N}\right)+6r_{V1}^r\Bigg]\,,
\ea
and
\ba
\label{cpivchpt}
c_V^\pi&=&
\frac{x_2}{60N}+x_2^2\Bigg[\frac{1}{12}k_1-\frac{1}{24}k_2+\frac{1}{24}k_6
+\frac{1}{3N}
\left(l_1^r-\frac{1}{2}l_2^r+\frac{1}{10}l_4^r+\frac{1}{2}l_6^r\right)
\nonumber\\&&
+\frac{1}{N}\left(-\frac{13}{540}L+\frac{1}{720}
-\frac{8429}{25920N}\right)+ r_{V2}^r
\Bigg]\,.
\ea

\subsection{Comparison with earlier work}

In addition to the previously mentioned checks, we can compare to earlier
partial results already available in the literature.

For the case of the vector form factor
we have checked that the dispersive part agrees up to a polynomial piece
with the analytical result for the dispersive part of Ref. \cite{tau}.
In addition, we have checked that the chiral logarithms that could be
obtained from chiral limit arguments agree with those given in
Ref. \cite{GM}. 

For the scalar form factor we agree with the earlier result for the pion mass
via the Feynman--Hellman theorem for $F_S(0)$. We also agree with the chiral
logarithms that were obtained in Ref. \cite{GM} and, finally we have
also checked that our expression for $F_S(s)$ has the correct absorptive
parts as derived in \cite{GM}.

\section{Resonance and $SU(3)$ estimates of the ${\cal O}(p^6)$ parameters}
\label{resonance}

To estimate higher order corrections due to scalar and vector
resonances in both form factors we follow in the remainder Ref.
\cite{EGPR}, and refer to it for the notation.
The SU(3) contributions come through kaon and eta
intermediate states. These estimates are of course scale dependent,
resulting therefore in a scale dependent final result.
We postpone the scale dependence study to Sect. \ref{numerics}.

The contribution of a given resonance state to $r_i^r$ is written
as $r_i^R$ with $R=S,V,K$ (scalars, vectors and kaons respectively),
since the effect of higher resonances is expected to be small
\cite{EGPR}. 

In this section, and the for the sake of simplicity,
we use a 2 by 2 matrix notation with
\ba
u_\mu = i u^\dagger \nabla_\mu\overline{U} u^\dagger =
u^\dagger_\mu,\nonumber\\ 
u^2 = \overline{U} = U^0-i\tau^i U^i\,,
\ea
the $\tau^i$ being the Pauli matrices and
$\nabla_\mu$ the relevant covariant
derivative as defined in Ref. \cite{EGPR}. We also use
\be
\chi_\pm = u^\dagger\chi u^\dagger\pm u\chi^\dagger u\,,\quad
f_{+\mu\nu} = u^\dagger (v_{\mu\nu}+a_{\mu\nu}) u +
u (v_{\mu\nu}-a_{\mu\nu}) u^\dagger
\ee
and $\langle A\rangle = \mbox{tr}(A)$.

\subsection{Scalar contributions}

In the scalar form factor the scalar resonance contribution is introduced
via the lagrangian 
\be
\label{LS}
{\cal L}[S(0^{++})] = \frac{1}{2}\langle\nabla_\mu S\nabla^\mu S-M_S^2 S^2
\rangle
+c_d \langle S u_\mu u^\mu \rangle
+c_m \langle S \chi_+ \rangle\,,
\ee
where $S$ contains the triplet and the singlet scalar in the
leading $1/N_c$ approximation.
This corresponds to use $\tilde{c}_m =c_m/\sqrt{3}$
and $\tilde{c}_d=c_d/\sqrt{3}$ in the notation of Ref. \cite{EGPR}.
We will use the numerical values
\be
M_S\approx 980 ~ \mbox{MeV}\;,\quad
c_m\approx 42 ~\mbox{MeV} \quad\mbox{and}\quad c_d \approx 32~\mbox{MeV}\,.
\ee
There are more possible terms than the ones we quote in  Eq.(\ref{LS}),
but there is not enough experimental information on the scalars to
determine them. 
Integrating out the scalars leads to the ${\cal O}(p^6)$ lagrangian
\be
\label{LS6}
{\cal L}^S=\frac{-1}{2M_S^4}
\left\{c_m^2\langle \chi_+ \nabla^2\chi_+\rangle
 + c_d^2\langle\nabla^2(u_\alpha u^\alpha)u_\mu u^\mu\rangle
 + 2 c_m c_d \langle u_\mu u^\mu\nabla^2\chi_+\rangle\right\}\,. 
\ee
The ${\cal O}(p^4)$ lagrangian also produced in this way is included here
via the contributions from $l_i^r$ terms.
There are obviously no contributions to $r_F^r$, $r_M^r$, $r_{V1}^r$
and $r_{V2}^r$ from Eq. (\ref{LS6}). For the rest we obtain
(using $F=F_\pi=93.2\,MeV$)
\ba
r_{S1}^S&=&0\;,\nonumber\\
r_{S2}^S&=&\frac{4F^2}{M_S^4} c_m\left(c_m-2 c_d\right)\approx -0.3~10^{-4}\;,
\nonumber\\
r_{S3}^S&=&\frac{4F^2}{M_S^4}c_m c_d\approx 0.5~10^{-4}\,.
\label{rSS}
\ea
These results should be taken as nothing more than an order of magnitude
estimate for the size of these constants.

\subsection{Vector contributions}

Similarly as was done in \cite{Buergi,BCEGS2,BT,BGS}, we use 
the formalism where the vector contribution
to the chiral Lagrangian starts at ${\cal O}(p^6)$.
The relevant lagrangian reads 
\be
{\cal L}[V(1^{++})] 
= \frac{-ig_V}{2\sqrt{2}} \langle \hat{V}_{\mu \nu}[u^\mu,u^\nu]\rangle
+ f_\chi\langle \hat{V}_\mu [u^\mu,\chi_-] \rangle-\frac{f_V}{2\sqrt{2}}
\langle \hat{V}_{\mu \nu} f_+^{\mu \nu}\rangle\,,
\ee
with $\hat{V}_\mu$ describing the vector meson and $\hat{V}_{\mu\nu}
=\nabla_\mu\hat{V}_\nu-\nabla_\nu\hat{V}_\mu$.
The parameter $f_V$ can be determined from $\rho\to e^+ e^-$\, \cite{EGLPR}
and $g_V$ and $f_\chi$ from $\rho\to\pi\pi$ and $K^*\to K\pi$\, \cite{BCEGS2},
(for the latter process we must use the extension to the three--flavour case).
This leads to 
\be
g_V = 0.09,\quad f_\chi=-0.03\quad\mbox{and}\quad f_V = 0.20\,.
\ee
We can now integrate out the vector meson degree of freedom
obtaining only a nonvanishing contribution for
\ba
r_{V1}^V& = &\frac{2\sqrt{2}f_\chi f_V F^2}{M_V^2}\;\approx\;-0.25~10^{-3}\,,
\nonumber\\
r_{V2}^V& = &\frac{g_V f_V F^2}{M_V^2}\;\approx\;0.26~10^{-3}\,,
\label{rVV}
\ea
where we have used $M_V=770$~MeV.

If instead we make use of full VMD in the three--flavour case, the form
factor would be described by
\be
F_V(q^2) = \frac{M_\rho^2}{M_\rho^2-q^2}
\approx 1 + \frac{q^2}{M_V^2} + \frac{q^4}{M_V^4}
- q^2\frac{ (M_\rho^2-M_V^2)}{M_V^4}+\ldots\,,
\ee
where the last term accounts for isospin--breaking effects.
We can now use as a simple estimate\footnote{A more thorough treatment
of quark mass corrections as done in \cite{BGT}
for the masses and decay constants would not change
any conclusions within the precision needed here.}
 $M_\rho^2-M_V^2=(M_{K^*}^2-M_\rho^2) M_\pi^2/M_K^2$, leading to very similar
numerical values for $r_{V1}^r$ and $r^r_{V2}$
\be
r_{V1}^r \approx -0.2~10^{-3}\quad\mbox{and}\quad r_{V2}^r
\approx 0.21~10^{-3}\,.
\ee

\subsection{$SU(3)$ contributions}

The SU(3) contributions to the low energy constants $r_i^r$ come 
from kaons and etas intermediate states.
For the vector form factor we take
the expression derived within $SU(3)$ CHPT \cite{GL3,BC}
and expand it in terms of  
inverse powers of $M_K^2$. This leads to $r_{V1}^K=0$ and
\be
\label{rv2K}
r_{V2}^K = \frac{F^2}{1920\pi^2 M_K^2} \approx 0.2~10^{-5}\,.
\ee
The result of Eq. (\ref{rv2K}) is trustable but
the one for $r_{V1}^K$ has several additional contributions coming from the 
relation between $L_9^r$ and $l_6^r$, the chiral logarithms and the
relations between $F_0$ and $F$ (the decay constant in the chiral limit
and in the limit with $m_u=m_d=0\ne m_s$ respectively \cite{GL2}).
We neglect the latter effects since for $\mu\approx 770$~MeV the derivative 
of $M_K^2\log(M_K^2/\mu^2)$ is very small.

In the scalar form factor case we can similarly expand the expressions
of $SU(3)$ CHPT as given in Ref. \cite{GL3} and obtain with
$M_K=495$~MeV
\ba
r^K_{S1}  &=&  0\,,
\nonumber\\
r^K_{S2} &=& \frac{F^2}{1152\pi^2M_K^2}\approx 0.3~10^{-5}\;,
\nonumber\\
r^K_{S3} &=& \frac{F^2}{384\pi^2M_K^2}\approx 0.9~10^{-5}\,.
\label{rKS}
\ea
$r^K_{S1}$ and $r^K_{S2}$ get contributions similar to those
discussed above and we also neglect them here. $r_F^K$ and $r_M^K$ only
get contributions of that type, so we set both to zero.

\section{Numerical results and comparison with experiment}
\label{numerics}

\subsection{General input parameters}
\label{input}

We use as input parameters $F_\pi=93.2$~MeV and
$M_\pi=M_{\pi^+}= 139.57$~MeV. We make also use of the
more commonly used
$\bar{l}_i$ quantities as defined in Eq. (\ref{defvarious}).
We will mainly use the following two sets of values 
\ba
\label{defsets}
\bar{l}_1 = -1.7,\quad\bar{l}_2 = 6.1,\quad\bar{l}_3 = 2.9&&\mbox{set I}\,,
\nonumber\\
\bar{l}_1 = -1.5,\quad\bar{l}_2 = 4.5,\quad\bar{l}_3 = 2.9&&\mbox{set II}\,.
\ea
The value of $\bar{l}_3$ is in both cases the one derived in \cite{GL1}.
Set I has the other two parameters obtained from the absolute
values of the $K_{l4}$ form factors using a dispersive improved
${\cal O}(p^4)$ CHPT calculation for three flavours \cite{BCG}.
Set II corresponds to using the ${\cal O}(p^6)$ 
calculation of $\pi\pi$ scattering
and fitting the $D$--wave scattering lenghts. This set also agrees,
within the errors, with other determinations of $\bar{l}_1$ and
$\bar{l}_2$ from dispersive analyses of $\pi\pi$ scattering data,
see Ref. \cite{KMSF2}, and Sect. 5.2 of Ref. \cite{BCEGS2} for a
discussion.

In Ref. \cite{GL1} a value of $\bar{l}_4 = 4.3$ was also obtained using
large $N_c$ arguments and the measured value of $F_K/F_\pi$.

\subsection{The scalar Form Factor}

\TABLE{

\addtolength{\tabcolsep}{-0.5mm}
\begin{tabular}{|c|c|c|ccc|c|c|}
\hline
 &&&&&&& \\
& ${\cal O}(p^2)$ & ${\cal O}(p^4)$ & \multicolumn{3}{c|}{${\cal O}(p^6)$ 
set I}&
                      ${\cal O}(p^6)$ set II&$r_i^R$ \\ 
$\mu$   (GeV)         & &       & 0.5  & 0.77 & 1.0  & 0.77 &     \\
\hline
 &&&&&&& \\
$\langle r^2\rangle^\pi_S$ (fm$^2$)
                  &  0  & 0.548 & 0.016& 0.017& 0.025& 0.054& $-$0.004 \\
$c^\pi_S$ (GeV$^{-4}$) & 0& 5.93  & 3.41 & 3.79 & 3.55 & 2.52 &  0.8 \\
 &&&&&&& \\
\hline
 &&&&&&& \\
$F_S(0)/(2B)$         &1&$-$0.0341&0.0081&0.0086&0.0080&0.0009&  0  \\ 
$F_\pi/F$             &1&~0.0611 &0.0061&0.0061&0.0063&0.0075& 0 \\
$M_\pi^2/M^2$         &1&$-$0.0206&0.0024&0.0026&0.0023&$-$0.0003& 0 \\
 &&&&&&& \\
\hline
\end{tabular}

\caption{\label{tableFS}Various contributions to the
scalar radius $\langle r^2 \rangle^\pi_S$, $c^\pi_S$, 
$F_S(0)/(2B)$, $F_\pi/F$ and to $M_\pi^2/M^2$. 
They all use $\bar{l}_4=4.3$, $r_i^r(\mu)=0$ for the columns labelled
${\cal O}(p^6)$. 
The quantities shown here do not depend on  $\bar{l}_{1,2}$ at
${\cal O}(p^4)$, and therefore are not sensitive to the use of set I or II
at this order. } 
}

Although the scalar form factor cannot be accessed experimentally, indirect
experimental information can be obtained from the data on $\pi\pi/KK$
scattering using dispersion relations \cite{DGL}: modulo an overall
normalization one can in fact derive the whole energy dependence of the
pion scalar form factor, since it goes to zero for
$s\to\infty$ as can be proven in perturbative QCD. 
The analysis of \cite{DGL} can be used to determine rather precise values
of $\langle r^2\rangle^\pi_S$ and $c^\pi_S$. 
The results for various parametrizations of the
relevant phase shifts, labelled by $A_1$, $A_2$ and $B$, give an
indication of the uncertainty involved in such quantities. The results are
-- we refer to \cite{GM} for further explanations --
\be
\label{valuesS}
\begin{array}{rclll}
\langle r^2\rangle^\pi_S&=&0.57\mbox{ fm}^2 (A_1)\;;& 0.61\mbox{ fm}^2
(A_2)\;;& 
0.60\mbox{ fm}^2 (B),\\
c^\pi_S&=&10.0\mbox{ GeV}^{-4} (A_1)\;;&10.9\mbox{ GeV}^{-4} (A_2)\;;
&10.6\mbox{ GeV}^{-4} (B)\,.
\end{array}
\ee
Following  \cite{GM}
we will use B as central value and the range as an
indication of the experimental error. The values labelled $A_1$ and $A_2$
use the CERN--MUNICH phase shifts \cite{CM} and those labelled B the
phase shifts of Au et al. \cite{Au}.

As was already noticed in \cite{GM}, the one--loop prediction of Gasser
and Leutwyler for the scalar radius (using $\bar{l}_4=4.3 \pm 0.9$),
\be
\langle r^2 \rangle^\pi_S = {x_2 \over 16 \pi^2} \left(6 \bar{l}_4 -{13
    \over 2} \right) 
+O(x_2^2) = (0.55 \pm 0.15) \mbox{ fm}^2 \; ,
\label{rs2_num_1}
\ee
is in very nice agreement with the dispersive evaluation of
Ref. \cite{DGL}.
Using the same value for $\bar{l}_4$ and set I for the other
constants, we can evaluate the corrections to the leading order result in
Eq. (\ref{rs2_num_1}) that come from the two--loop calculation (here we use 
also $\mu=M_\rho$, and $r^r_{S2}(M_\rho)=0$) 
\be 
\langle r^2 \rangle^\pi_S = 0.548 + 0.017 = 0.565 \;\mbox{fm}^2 \; .
\label{rs2_num_2}
\ee
These are rather small and go in the right direction to
improve the agreement with the experimental information
Eq. (\ref{valuesS}). This shows that for this quantity the 
convergence of the chiral expansion is quite fast, and hence that it can 
be used reliably to determine rather precisely the constant $\bar{l}_4$.
The central value $\langle r^2 \rangle^\pi_S = 0.60$~fm$^2$ given by
solution B, is exactly reproduced by $\bar{l}_4= 4.47\; (4.29)$, if we use
set I (set II) together with $\mu=M_\rho$ and
$r^r_{S2}(M_\rho)=0$. The influence of the latter constant on the value of
$\bar{l}_4$ is tiny, and can be neglected altogether. There is some
dependence on the choice of the scale, as it is illustrated by a variation
of $\mu$ between 1 and 0.5 GeV: $\bar{l}_4$ varies between 4.43 and 4.47
for set I, and 4.21 and 4.35 for set II. Taking into account the
uncertainty in the determination of $\langle r^2 \rangle^\pi_S$, and
allowing for a range of values between 0.57 and 0.63 fm$^2$, we can
conclude that
\be 
\bar{l}_4 = 4.4 \pm 0.3 \; \; ,
\ee
after averaging over what would be obtained with either set I or
II (in principle, once the values of $\bar{l}_{1,2}$ will be better
determined, the error on $\bar{l}_4$ could be reduced even more). We stress
that, given the good convergence of the chiral expansion in 
this case, the effect of yet higher orders (beyond two loops) can be safely
neglected. 

\FIGURE{
\epsfxsize=12cm\epsfbox{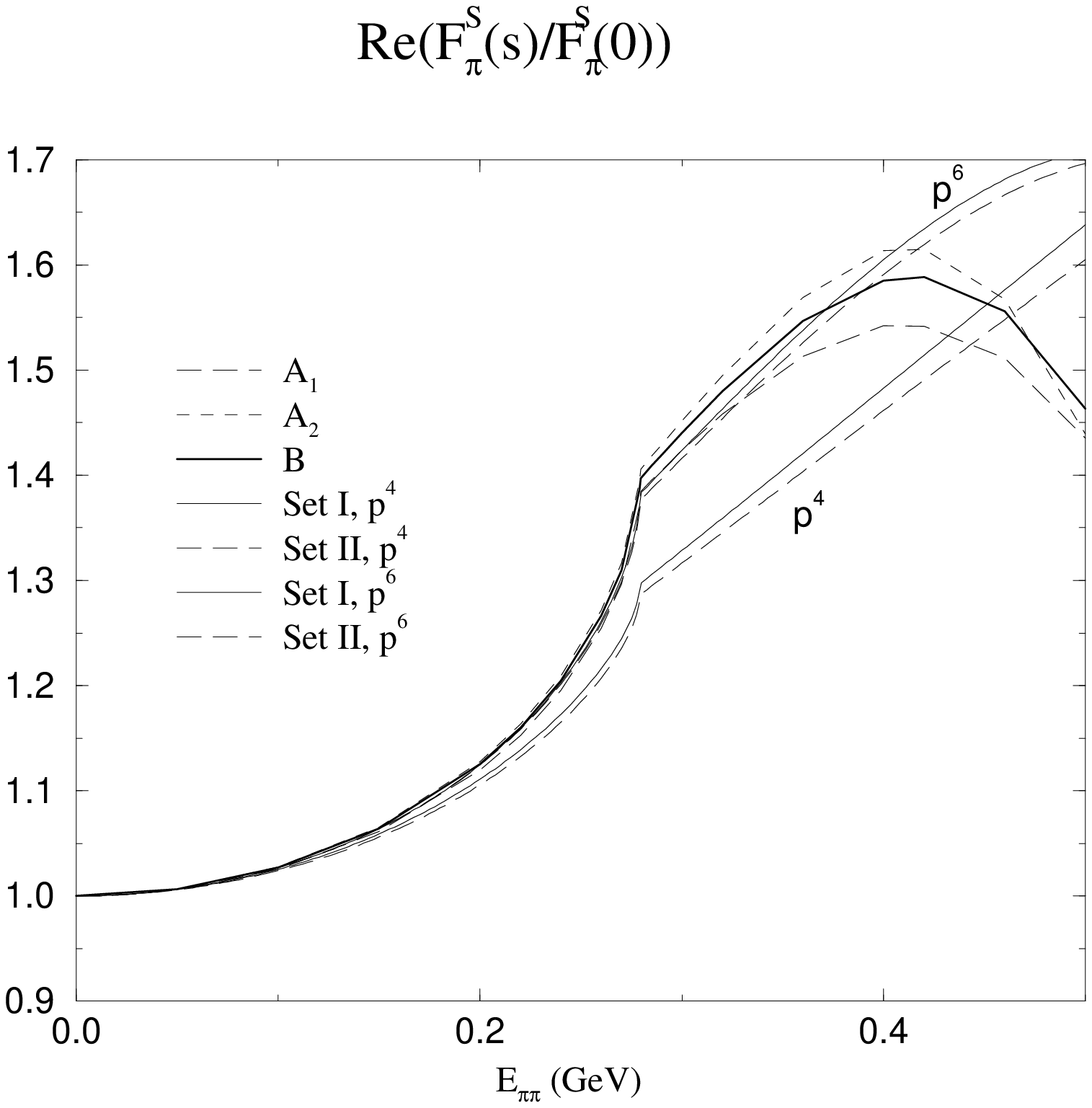}
\caption{\label{plotFS} The real part of the normalized scalar form factor
for the two sets of parameters that reproduce the scalar radius and,
$c_S^\pi$ of set $B$, for the two--loop case (${\cal O}(p^6)$) and the
one--loop case (${\cal O}(p^4)$). 
 }}

The value of the coefficient $c_S^\pi$ at one loop does not depend on low
energy constants, but it turns out to be quite far from
the experimental number
\be
c_S^\pi = {x_2 \over 16 \pi^2} {19 \over 120} = 5.93 \mbox{ GeV}^{-4} \; \;
{}. 
\ee
At two loops the situation improves considerably, although the exact value
of this coefficient now depends both on ${\cal O}(p^4)$ and 
${\cal O}(p^6)$ low energy constants. Using the value of $\bar{l}_4$
determined with the scalar radius, we get the following numerical values
\ba
c_S^\pi (\mbox{set I}) &=& 9.85~\mbox{GeV}^{-4}+x_2^2 r_{S3}^r(M_\rho) \;
\; , \nonumber \\
c_S^\pi (\mbox{set II})&=& 8.59~\mbox{GeV}^{-4}+x_2^2 r_{S3}^r(M_\rho) \;
\; .
\ea

\FIGURE[tbh]{
\epsfxsize=12cm\epsfbox{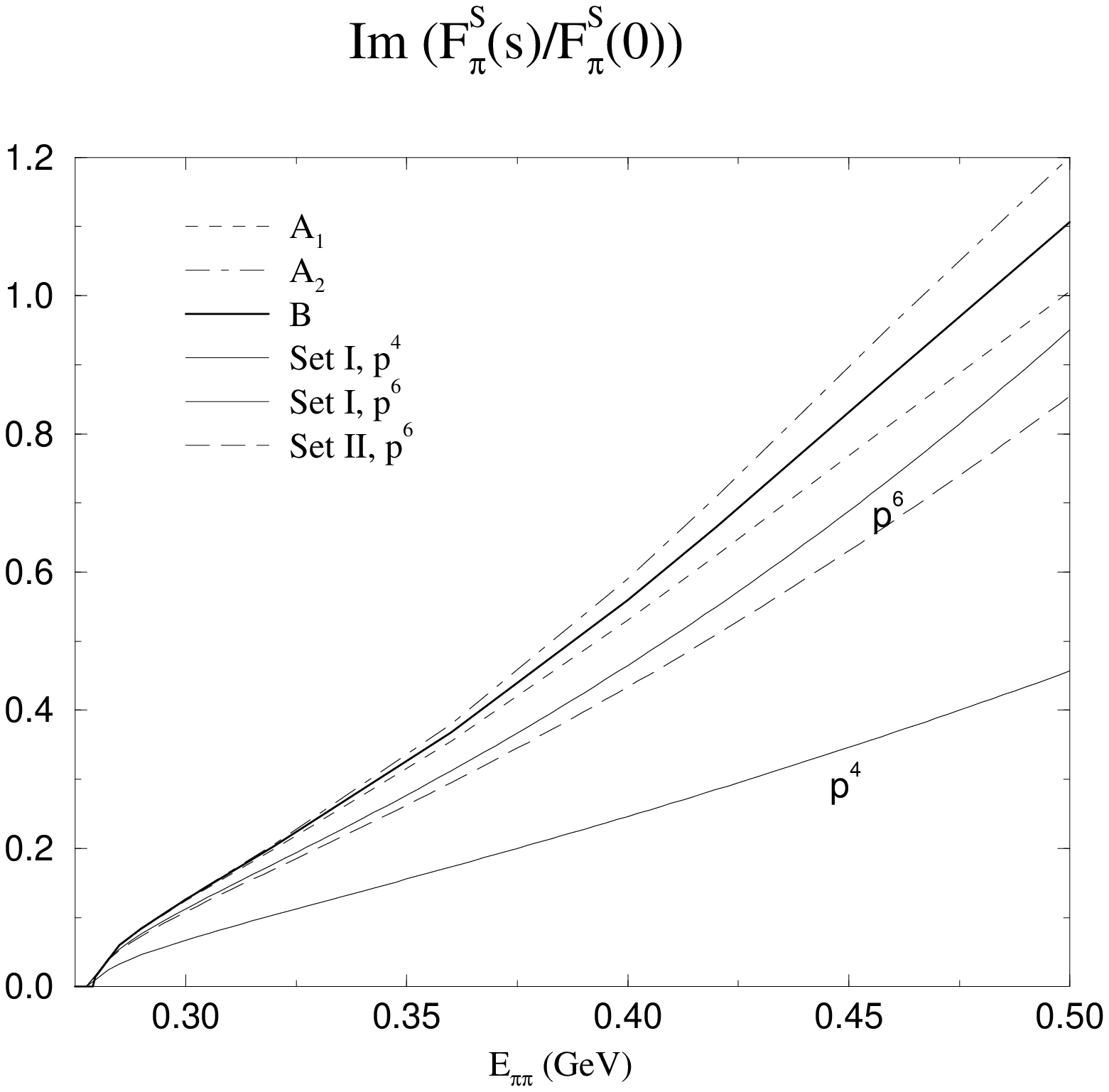}
\caption{\label{plotFS2}The imaginary part of the normalized scalar
form factor for the two sets of parameters that reproduce
the scalar radius, and $c_S^\pi$, of set $B$ for the two--loop case 
(${\cal O}(p^6)$)
and the one--loop case (${\cal O}(p^4)$), which is identical for both.}
}

The two--loop correction to the leading order result is therefore of the
right size needed to bring the theoretical number very close to the
dispersive determination. To get perfect agreement we could just tune
$r_{S3}^r(M_\rho)$ accordingly, and get $r_{S3}^r(M_\rho)=5.6 \times
10^{-5} (1.5 \times 10^{-4})$ for set I (set II). This fine tuning of the
${\cal O}(p^6)$ constant is not especially interesting, were it not for the
fact that the value we get is rather close to what we obtained with the
resonance saturation hypothesis, Eq. (\ref{rSS}, \ref{rKS}). Notice that
the value of $r_{S3}^r(M_\rho)$ varies by a factor of three when using set I
or II, and we should be rather satisfied with an order of magnitude and
sign agreement. 

If we simply assume naive scalar dominance for
the $l_i^r$ and $r_i^r$ contributions to the scalar form factor
\be
\label{naiveS}
1 + {1 \over 6} \langle r^2\rangle^\pi_S s_\pi + c^\pi_S s_\pi^2
\approx
\frac{M_S^2}{M_S^2-s_\pi}\,,
\ee
we would have obtained $r_{S3}^r=8.2~10^{-5}$
and $l_4^r=9.0~10^{-3}$ using a value of $M_S=980$~MeV.
Notice that these values are not so far from the observed ones, being
of the right order of magnitude\footnote{A value of
$\bar{l}_4=4.4$ corresponds to $l_4^r=6.2~10^{-3}$
at $\mu=0.77$~GeV.}. A scalar mass of 500~MeV
would have increased the values of $l_4^4$ and $r_{S3}^r$
by a factor of 3.8 and 15 respectively,
bringing them far out of the region determined from experiment.

In Fig. \ref{plotFS} we have shown the
real part of the scalar form factor
at ${\cal O}(p^4)$ and at ${\cal O}(p^6)$ 
for set I, $\mu=0.77$~GeV, $\bar{l}_4=4.472$,
$r_{S2}^r=0$ and $r_{S3}^r=4.9~10^{-5}$,
as well as for set II, $\mu=0.77$~GeV, $\bar{l}_4=4.29$,
$r_{S2}^r=0$ and $r_{S3}^r=16.4~10^{-5}$.
In Fig. \ref{plotFS2} we have shown the imaginary part
for the same approximations, which is of course
identical for the two sets at ${\cal O}(p^4)$.

\subsubsection{$F_S(0)$, $F_\pi/F$ and $M_\pi^2/M^2$}

In table \ref{tableFS} we also show the various contributions to
the value of the scalar form factor at zero momentum transfer compared to
its lowest order value. As can be seen, the ${\cal O}(p^6)$ correction
is quite small here. For set I, $\mu=0.77$~GeV and $\bar{l}_4=4.4$
the value is $F_S(0)/(2B)=0.974+x_2^2 r_{S1}^r$. $F_S(0)$ is of course
most sensitive to the value of $\bar{l}_3$ on which we have no extra
information here.
For completeness we have also included in Table \ref{tableFS}
the corrections to the ratios $F_\pi/F$ and $M_\pi^2/M^2$.
If we use the value $\bar{l}_4=4.4$ calculated assuming the value
of $r_{S2}^r$ derived from scalar exchange we obtain instead
\be
\frac{F_\pi}{F} = 1.069\pm0.004
\quad\mbox{and}\quad
\frac{M_\pi^2}{M^2} = 0.982\,.
\ee
The error is determined by looking at the variation in table \ref{tableFS}.
We do not quote an error on $M_\pi^2/M^2$ since we have no improved
information on $\bar{l}_3$ here.

\subsubsection{Modified Omn\`es representation.}

The unitarity condition which must be obeyed by the scalar form factor is
satisfied by the following explicit representation, which is due to
Omn\`es:
\begin{equation}
F_S(s) = P(s) e^{\Delta_0(s)} \; \; ,
\label{omnes}
\end{equation}
where
\begin{equation}
\Delta_0(s) = {s \over \pi} \int_{4 M_\pi^2}^{\infty} {ds' \over s'}
  {\phi(s') \over s'-s-i\epsilon} \; \; ,
\end{equation}
and $P(s)$ is a polynomial which, in the case of the scalar form
factor, can be taken to be a constant. In principle, if one would know the
phase and inelasticity of $\pi \pi$ for $I=0$, $S$--wave, one would know also
the scalar form factor up to a constant.
Since CHPT provides a representation for the phase $\delta_0^0$, one could use
Eq. (\ref{omnes}) to exponentiate the result obtained at any given order of
the chiral expansion. As discussed in \cite{GM},
however, there are problems in carrying through this procedure,
particularly because of the bad high--energy behaviour of the chiral
representation for the phase shifts. 

The authors of \cite{GM} have instead proposed what they called the ``Modified
Omn\`es representation''(MOR), which is defined in the following manner.
First they defined the reduced form factor $\Gamma^\Lambda(s)$, as
\begin{equation}
\Gamma^{\mbox{\tiny{CHPT}}} = e^{\Delta_\Lambda(s)} \Gamma^\Lambda(s) \; \; ,
\label{gamma_l}
\end{equation}
with
\begin{equation}
\Delta_\Lambda(s) = {s \over \pi} \int_{4 M_\pi^2}^{\Lambda^2} {ds' \over s'}
{\phi(s') \over s'-s-i\epsilon} \; \; .
\end{equation}
The reduced form factor has the following analytic properties:
\begin{enumerate}
\item
it is analytic in the complex plane, except for a cut along the positive real
axis starting at $s=16 M_\pi^2$;
\item
it satisfies the dispersion relation
\begin{equation}
\Gamma^\Lambda(s) = 1 + {s \over \pi} \int_{16 M_\pi^2}^{\infty} {ds' \over
  s'} {\mbox{Im} \Gamma^\Lambda(s') \over s'-s-i\epsilon} \; \; ;
\end{equation}
\item in the region $16 M_\pi^2<s<M_K^2$ Im$\Gamma^\Lambda(s)$ only gets
contributions from many particle intermediate states and is hence small.
\end{enumerate}
Given these properties it is easy to show that at
low energy $\Gamma^\Lambda(s)$
can be well approximated by a polynomial. Inserting a polynomial of a given
order for $\Gamma^\Lambda(s)$ on the right--hand side of Eq. (\ref{gamma_l}),
one obtains the Modified Omn\`es representation.

Gasser and Mei{\ss}ner have then compared the MOR which they obtained using a
linear polynomial for $\Gamma^\Lambda$ and the phase shift to one loop. Since
the CHPT phase is now known to two loops, we can check here what kind of
improvements this yields for the MOR. Besides this we also use a quadratic
polynomial for $\Gamma^\Lambda$, and fix its coefficients such that we
reproduce the Taylor expansion at $s=0$ as given by the exact solution.
The results are shown in Fig. \ref{fig:MOR}, where we compare the MOR with
one-- or two--loop phase shifts to the exact solution.
\FIGURE[tbh]{
\epsfxsize=11cm\epsfysize=11cm\epsfbox{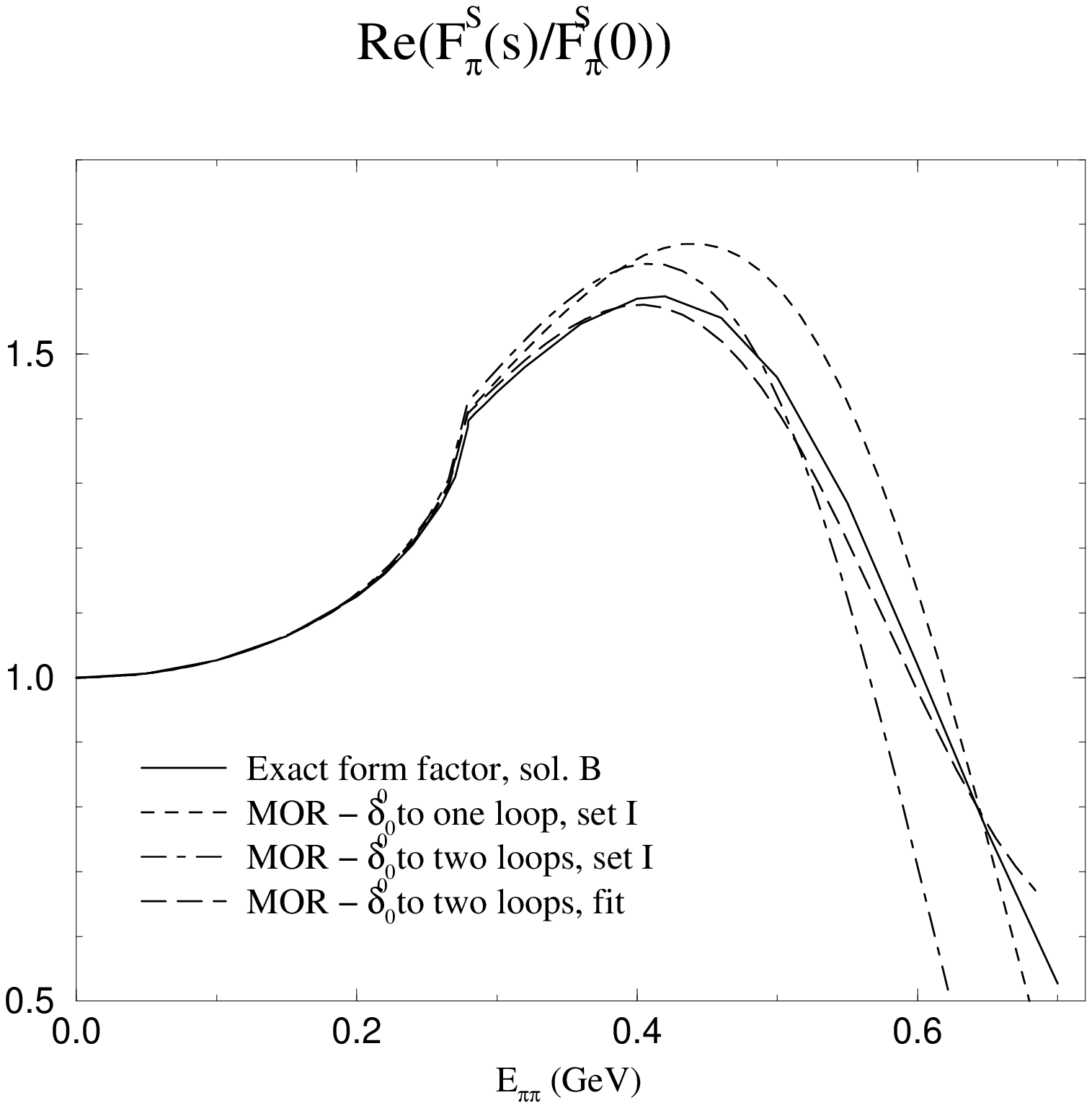}
\caption{\label{fig:MOR} Comparison of the Modified Omn\`es representation 
  (MOR) for the scalar form factor with the exact solution B. For the MOR
  we have three curves: two are made with set I for the LEC's, and one-- or 
  two--loop pion phase shift. The third one (identified by the word
  ``fit'') has been produced with the phase shift at two loops, and  by
  choosing ad hoc values for the constants $\bar{l}_1=-1.5, \;
  \bar{l}_2=5.0$, so as to reproduce as closely as possible the exact
  solution.} }
As can be seen from the figure, the qualitative features of this
representation do not change whether one uses the one-- or two--loop
expression for the phase. There is of course a quantitative change which,
however, is not very large up to 700 MeV. 
On this basis one could argue that with this representation and the
two--loop phase as input, one can get a rather good description of the form
factor up to $\sim 600$--700 MeV. 
The curves relative to set I, that are shown in the figure, seem to
contradict this statement as their comparison to the exact form factor is
not very good  already immediately after threshold. On the other hand the
situation improves drastically if we use set II for the LEC's and the
two--loop phase.
As can be seen from Fig. \ref{fig:MOR}, an impressive agreement up to 700
MeV can be obtained by changing $\bar{l}_2$ by half a unit with respect to
that of set II, i.e. $\bar{l}_2= 5.0$.
We do not want to emphasize too much this agreement, since we do not
discuss in detail the uncertainties involved. On the other hand we find it
interesting that also the scalar form factor analyzed in this manner seems
to indicate the need for a lower value of $\bar{l}_2$ than what was given
by the $K_{l4}$ analysis \cite{BCG}.
In addition, this example just shows very clearly how one can use unitarity
to improve the chiral representation, and push it somewhat beyond its
typical limits of validity.

\subsection{Comparison with data for $F_V$}

\subsubsection{The data and previous analyses}

The pion form factor has been measured both in the timelike, $s>0$,
and in the spacelike, $s<0$, region. In the spacelike region there
are two experiments with a reasonably large data set,
Dally et al. \cite{Dally} and NA7 at CERN \cite{Amendolia86}.
The latter set is an accurate measurement of $\pi e$ elastic scattering
and dominates all fits. It agrees with \cite{Dally} in the overlap
region, where it has significantly smaller errors.

In the timelike region there are more experiments but none of them has a
large and accurate data set in the region relevant here. A review of the
data before the recent inclusion of $\tau$--decays can be found in
Ref. \cite{Barkov}. 
The data are obtained in three ways: $\tau$--decays to $\pi\pi\nu_\tau$
\cite{ALEPH}, $e^+ e^-\to\pi^+\pi^-$ in electron positron colliders
\cite{Barkov,Barkov79,Quenzer,Wasserman} and $e^+ e^-\to\pi^+\pi^-$
measured in NA7 \cite{Amendolia84}.

One value of the pion charge radius was obtained in
\cite{Amendolia86} using a pole fit leaving the normalization free,
leading to the result\footnote{See below for the definition of $n$}
\be
\rpiV = 0.431\pm 0.010\mbox{ fm}^2\quad n = 0.995\pm0.002\,.
\label{rVst}
\ee
They also used the parametrization of the vector form factor of
Ref. \cite{Dubnicka}, which is a Pad\'e approximation to the Omn\`es
formula using the $\delta_1^1$ $\pi\pi$ phase shift, and obtained
\be
\rpiV = 0.439\pm0.008~\mbox{fm}^2\,.
\ee
In \cite{BC} the same data were fitted to the one--loop CHPT formula
obtained in \cite{GL2} for the three--flavour case.
There, the values $\rpiV = 0.392$~fm$^2$ and $\rpiV = 0.366$~fm$^2$
were obtained from a fit with normalization one and free respectively.
The $\chi^2$ were not as good as for the pole fits and in the latter
case the normalization ended up outside the error band given in
\cite{Amendolia86}. The main cause was that the one--loop chiral
formula did not satisfactorily describe the higher $|s|$ data.

In Table \ref{fitFV1} we have shown the results of a fit to various sets
of data (as specified in the table and explained below
in subsection \ref{FV2CHPT}), using Eq. (\ref{FVexpansion}), and
also a pole formula like
\be
\label{FVinverse}
F_V(s) = \frac{1}{1-\frac{1}{6}\rpiV s}+\hat{c}_V^\pi s^2\,.
\ee
Notice that in this case 
\be
c_V^\pi={1 \over 36} \left(\langle r^2 \rangle ^\pi_V\right)^2 
 + \hat{c}_V^\pi \; \; . 
\ee
Only for the data of \cite{Amendolia86} we have taken a normalization
uncertainty into account. For none of the other experiments
is the systematic error even close to the statistical error, with
the possible exception of \cite{Amendolia84}. E.g. the CMD
data of \cite{Barkov} have less than a 2\% systematic uncertainty
at every point.
So in all fits below and in the next subsubsection we
multiply $F_V$ by $n$ for fitting data from \cite{Amendolia86}, and
by one for all the other data sets.
In the experiment $n = 1.0000\pm0.0045$.

\TABLE{

\begin{tabular}{|c|c|cccc|}
\hline
 &&&&& \\
Param.& data set & $\chi^2$/dof & $n$ & $\rpiV$ (fm$^2$) & $c_V^\pi$
(GeV$^{-4}$)\\ 
 &&&&& \\
\hline
 &&&&& \\
&\cite{Amendolia86} &
42.4/42 & $0.995\pm0.002$ & $0.420\pm0.019$ & $2.4\pm0.5$\\
Polynom&\cite{Amendolia86} (cut) &
17.8/22 & $1.000\pm0.005$ & $0.478\pm0.056$ & $5.1\pm2.7$\\
Eq.(\ref{FVexpansion})&Spacelike &
50.5/55 & $0.996\pm0.002$ & $0.429\pm0.016$ & $2.6\pm0.4$\\
&Timelike &
23.1/20 & irrelevant & $0.189\pm0.098$ & $10.6\pm2.3$\\
&All & 
87.7/77 & $0.999$ & $0.459\pm0.009$ & $3.5\pm0.2$\\
 &&&&& \\
\hline
\hline
 &&&&& \\
&\cite{Amendolia86} &
41.2/42 & $0.997\pm0.003$ & $0.442\pm0.022$ & $3.85\pm0.68$\\
Pole&\cite{Amendolia86} (cut)&
17.7/22 & $1.000\pm0.005$ & $0.488\pm0.059$& $6.18\pm3.1$ \\
Eq. (\ref{FVinverse})&Spacelike &
48.9/55& $0.997\pm0.002$&$0.447\pm0.018$ & $4.0\pm0.57$\\
&Timelike &
23.2/20& irrelevant & $0.191\pm0.105$ & $10.4\pm2.8$\\
&All & 
76.7/77 & $0.995\pm0.002$ & $0.427\pm0.007$ & $3.36\pm0.22$\\
 &&&&& \\
\hline
\hline
 &&&&& \\
&\cite{Amendolia86} &
41.8/42 & $0.996\pm0.002$ & $0.431\pm 0.019$ & $3.20\pm0.51$\\
CHPT&\cite{Amendolia86} (cut) &
17.8/22 & $1.000\pm0.005$ & $0.482\pm 0.056$ & $5.59\pm2.8$\\
Eq. (\ref{FV})&Spacelike &
49.7/55 & $0.996\pm0.002$ & $0.438\pm 0.016$ & $3.35\pm0.44$\\
&Timelike &
22.9/20 & irrelevant & $0.134\pm 0.098$ & $11.4\pm2.3$\\
&All & 
84.2/77 & $0.998\pm0.002$ & $0.448\pm0.009$ & $3.68\pm0.24$\\
 &&&&& \\
\hline
 &&&&& \\
Eq. (\ref{FV})&All & 
80.7/76 & $0.996\pm0.002$ & $0.437\pm0.011$ & $3.84\pm0.25$\\
+$d^\pi_V s^3$&All but \cite{Amendolia84} & 
58.2/72 & $0.997\pm0.002$ & $0.453\pm0.011$ & $4.45\pm0.28$\\
 &&&&& \\
\hline
\end{tabular}

\caption{\label{fitFV1} Various fits to the pion form factor data
using the simple parametrizations Eq. (\ref{FVexpansion}), for the first
five rows, the pole formula Eq. (\ref{FVinverse}) for the second five
rows, and the full two--loop CHPT expression, Eq. (\ref{FV}), for the
remaining seven. The last two fits include an extra parameter of the form
$d_V^\pi s^3$ added to the two--loop CHPT expression, which the fit
determines to be: $d_V^\pi = 3.0 \pm1.6 $GeV$^{-6}$ when all data are
fitted, and $d_V^\pi = 4.1 \pm1.6 $GeV$^{-6}$ when all data but those of
Ref. \cite{Amendolia84} are fitted. 
In the second of the five data sets used, we applied a cut
and used only those data satisfying $\sqrt{-s}<300$~MeV.
The errors are those that change the $\chi^2$ by one. All have a free
normalization for the data of \cite{Amendolia86}. See text for details.} 
}

\subsubsection{Comparison of $\langle r^2 \rangle ^\pi_V$ and $c_V^\pi$
  with  CHPT at two loops}
\label{FVCHPT}

The charge radius of the pion has been used by Gasser and Leutwyler to
determine the low energy constant $\bar{l}_6$ with the result: $\bar{l}_6 = 
16.5 \pm 0.9$, that reproduces $\langle r^2 \rangle ^\pi_V=0.439 \pm 0.03
\mbox{ fm}^2$. 
Since we do not have other sources of information on
$\bar{l}_6$, CHPT to two loops can only be used here to refine the
determination of this constant. 
It is instructive to rewrite the two--loop correction in the following
form 
\be
\label{rerV}
\langle r^2 \rangle^\pi_V = {x_2 \over N } \left[ \left( 1+2 {x_2 \over N}
    \bar{l}_4 \right) \left( \tilde{l}_6 - 1 \right) + {x_2 \over N}
    \left( N {13 \over 192} - {181 \over 48} \right) \right] \; \; ,
\ee
where we have defined a new constant
\be
\tilde{l}_6 = \bar{l}_6 + 6  x_2  \left[ N r^r_{V1} + {1 \over 
    3} L \left( {19 \over 12} -\bar{l}_1 +
    \bar{l}_2 \right) \right] \; \; ,
\ee
which differs from $\bar{l}_6$ by a scale--independent quantity.
{}From this expression it is clear that (besides the last piece in
Eq. (\ref{rerV}), which is a tiny effect) the two--loop correction to the
charge radius consists of two main contributions.
Part of the correction is due to the renormalization of $F \rightarrow
F_\pi$ in the leading term, and produces the factor $1+2 x_2/ N \bar{l}_4$,
which numerically represents a 12\% correction. 
The other part is a pure two--loop effect, and shifts $\bar{l}_6$ into 
$\tilde{l}_6$.
Numerically, this last effect is as follows (using set I)
\be
\tilde{l}_6-\bar{l}_6 = -0.91 + 6 N x_2 r^r_{V1}(M_\rho) = -1.44 \; 
\; ,
\ee
where the numerical value after the last equal sign has been obtained
inserting for $r^r_{V1}(M_\rho)$, our VMD estimate Eq. (\ref{rVV}).
(The use of set II shifts the above values by $+0.17$).
Modulo the uncertainty coming from the contribution of the ${\cal O}(p^6)$
LEC, this effect is of the order of $-10$\%.

The two main effects of the two--loop correction, 
both around ten percent level, contribute with opposite sign, and tend
to cancel each other, resulting in a rather small correction to the radius 
\ba
\langle r^2 \rangle^\pi_V (\mbox{set I}) &=& 0.440 + 0.032 + 60.3
\cdot r^r_{V1}(M_\rho) =  0.457 \mbox{~fm}^2 \; \; , \nonumber \\
\langle r^2 \rangle^\pi_V (\mbox{set II}) &=& 0.440 + 0.037 + 60.3
\cdot r^r_{V1}(M_\rho) = 0.462 \mbox{~fm}^2\; \; , 
\ea
where the final value has been obtained using our estimate for the constant 
$r^r_{V1}(M_\rho)$, Eq. (\ref{rVV}). This result shows that also here, the
chiral expansion is converging rapidly, and that the determination
of $\bar{l}_6$ through the charge radius is in principle rather reliable.
On the other hand, in this case the new LEC at ${\cal O}(p^6)$ plays a more
important role than the corresponding one for the scalar radius case, where
the size of the new LEC suggested by resonance saturation gave a negligible
contribution. 
It is clear that lacking an independent source of information on
$r^r_{V1}(M_\rho)$, we can use the data on the charge radius only to
determine the constant $\tilde{l}_6$.
Any attempt to determine only $\bar{l}_6$ will depend on
estimates and/or assumptions on the value of $r^r_{V1}(M_\rho)$, at least
at the level of a $\pm 10$\% uncertainty. 

In the case of $c_V^\pi$, the contribution of the ${\cal O}(p^6)$ LEC
$r^r_{V2}(M_\rho)$ is even more important
\be
c^\pi_V = 0.62 + 1.96 + 1.3 \times 10^4 r^r_{V2}(M_\rho) = 
 5.4 \mbox{~GeV}^{-4}  \; \; ,
\ee
where the first factor refers to the one--loop contribution, and the second
to the two--loop one with $r^r_{V2}(M_\rho)=0$, evaluated using the old
value $\bar{l}_6=16.5$ and set I, and where the final number has been
obtained using our VMD estimate Eq. (\ref{rVV}). The coefficient $c^\pi_V$
is therefore mainly sensitive to the value of $r^r_{V2}(M_\rho)$, and in
principle can be used to determine this ${\cal O}(p^6)$ LEC with rather small
uncertainties (modulo higher--order contributions). 

\subsubsection{Fit to the data with CHPT at two loops}
\label{FV2CHPT}
Besides comparing the vector form factor 
CHPT formulae with the ``experimental values'' of the Taylor expansion 
coefficients at $s=0$, we can attempt a more ambitious use of our two--loop
results, and directly fit the data. 
The reason for this is twofold: first, there are abundant and accurate data
precisely in the region of energy where CHPT can be trusted more. Second,
in that region, CHPT is certainly less model dependent than other 
parametrizations used to fit the data (see above), that make various kind
of assumptions. The only assumption made in using a CHPT expression is in
the truncation of the expansion to a given order -- in the present case, the
only theoretical bias comes from neglecting contributions of three--loop
order in the chiral expansion. As we will see, we can easily estimate such
higher order effects and therefore obtain a reliable model--independent
value for the two parameters $\langle r^2 \rangle_V^\pi$ and $c_V^\pi$. 

Our fit has been made using the expression:
\begin{equation}
\label{fvchiral}
F_V(s) =  1 + \frac{1}{6}\langle r^2 \rangle^\pi_V s + c^\pi_V s^2 
          + f_V(s)
          + O(s^3)\quad,
\end{equation}
where $f_V(s)$ can be easily obtained from Eqs. (\ref{FV},\ref{UV}). As
free parameters in the fit we have used $\langle r^2 \rangle^\pi_V s$ and
$c^\pi_V$. It is clear that the advantage of CHPT is that it provides a way 
to calculate explicitly the function $f_V(s)$, whereas in all other cases
one can only make guesses about its value.  We remark that although the
exact numerical value of this function depends on the LEC's, their effect
is rather small, and the uncertainty in the knowledge of their value can be 
neglected altogether.
The results of our fits are presented in Table \ref{fitFV1}, where we have
fitted, as before, five different data sets, as specified in the Table. 
As can be seen there, the results of the CHPT fits are rather close to both
those obtained with the simple polynomial parametrization,
Eq. (\ref{FVexpansion}), and to those obtained with the pole formula,
Eq. (\ref{FVinverse}). 
This shows that the assumptions made in the two different
parametrizations, Eq. (\ref{FVexpansion}) and Eq. (\ref{FVinverse}), are
reasonable, and can in fact be partly justified on the basis of CHPT. 

To estimate yet higher orders in CHPT we have chosen to introduce an extra
term
in the polynomial part of Eq. (\ref{fvchiral}), of the form $d_V^\pi
s^3$, and to fit this new parameter from the data. The value we find is
$d_V^\pi = 3.0 \pm 1.6$ GeV$^{-6}$ when we fit all the data, and $d_V^\pi =
4.1 \pm 1.6$ GeV$^{-6}$ when we fit all the data except those of
Ref. \cite{Amendolia84}. 
The changes in the values of the charge radius, and of $c_V^\pi$, are rather 
small, as can be seen in Table \ref{fitFV1}.
Moreover, the improvement in the $\chi^2$ is visible but not dramatic.
This shows that there is no need for a large contribution from higher
chiral orders, and that we can confidently use this extra parameter
$d_V^\pi$ to estimate their effect.
All in all, we conclude that the best values for the charge radius and
$c_V^\pi$ are given by
\ba
\rpiV &=& (0.437\pm0.016)\mbox{ fm}^2\; ,
\nonumber\\
c_V^\pi &=& (3.85\pm0.60)\mbox{ GeV}^{-4} \; ,
\label{rpiVnumber}
\ea
where the error also takes into account the theoretical uncertainty
(i.e. we have added in quadrature the statistical error coming from the
fit, and the difference in the central values between the fits with and
without the cubic term in the polynomial -- for the error on $c_V^\pi$,
however, see below).

In Fig. \ref{figFVspace} we have plotted the available experimental data in
the spacelike region together with the results of the two fits
corresponding to the last two rows of Table \ref{fitFV1}. The curves
corresponding to the same two fits have been plotted in the timelike
region, together with the experimental data, in Fig. \ref{timevector}. 
Notice that the $\chi^2$ improves significantly 
if the timelike NA7 data \cite{Amendolia84} is not considered. Our believe
is that in the latter experiment the systematic errors in the timelike
region are underestimated. For this reason we have made a fit including
all data but those of Ref. \cite{Amendolia84}, to show how much the central 
values would be shifted. The shift in the charge radius would still be
within the error bars, while that for $c_V^\pi$ not: in view of this we
have enlarged the error bar of this quantity accordingly (see
Eq. (\ref{rpiVnumber})). 

\FIGURE[tbh]{
\epsfxsize=14cm\epsfbox{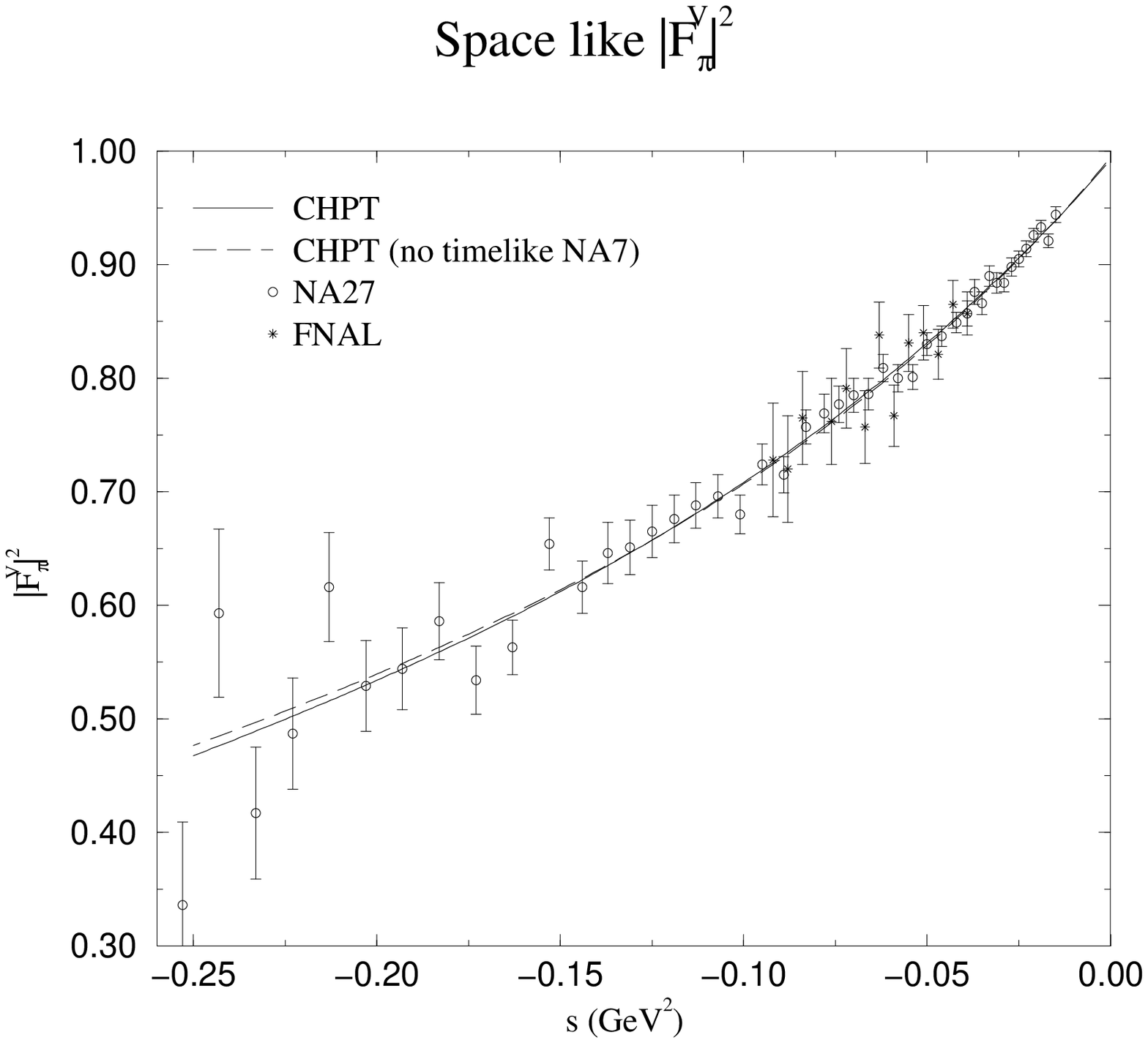}
\caption{\label{figFVspace} The space like data for the vector form factor
  and the curves corresponding to the fits in the last two rows of Table
  \ref{fitFV1}, i.e. using an expression like Eq. (\ref{fvchiral}) with the 
  addition of a cubic term in the polynomial.} }

We can now use the above charge radius and $c_V^\pi$
determinations to extract the values of the CHPT LEC's 
$\bar{l}_6$ and $r^r_{V2}$.
As we have already stressed, one can unambiguously determine from the charge
radius only the constant $\tilde{l}_6$
\be
\tilde{l}_6=14.6 \pm 0.5 \; \; ,
\ee
and if we want to translate this into a value for $\bar{l}_6$, we have to
make an estimate on the value of $r^r_{V1}$. Using our VMD estimate,
Eq. (\ref{rVV}), we get
\be
\bar{l}_6=16.0 \pm 0.5 \pm 0.7 \; \; ,
\ee
where the last error is purely theoretical, and takes into account the
uncertainty due to the $r^r_{V1}$ estimate (which we assume to be
$\pm 100$\%), and the uncertainty in our knowledge of $\bar{l}_{1,2}$,
which also enter $\tilde{l}_6$.
Compared to the one--loop determination of the same constant, made in
\cite{GL1}, we have found a very similar central value
(due to the smallness of the two--loop correction we have calculated), but
we are able now to make a more reliable error estimate. As we have
discussed, there is no way to reduce the error indicated here, which is
mainly theoretical. 
Finally, from the $c_V^\pi$ value, Eq. (\ref{rpiVnumber}), we get
\be
r^r_{V2}(M_\rho) = (1.6 \pm 0.5) \cdot 10^{-4} \; \; ,
\ee
which is in reasonable agreement with our VMD estimate Eq. (\ref{rVV}).
\FIGURE[tbh]{
\epsfxsize=14cm\epsfbox{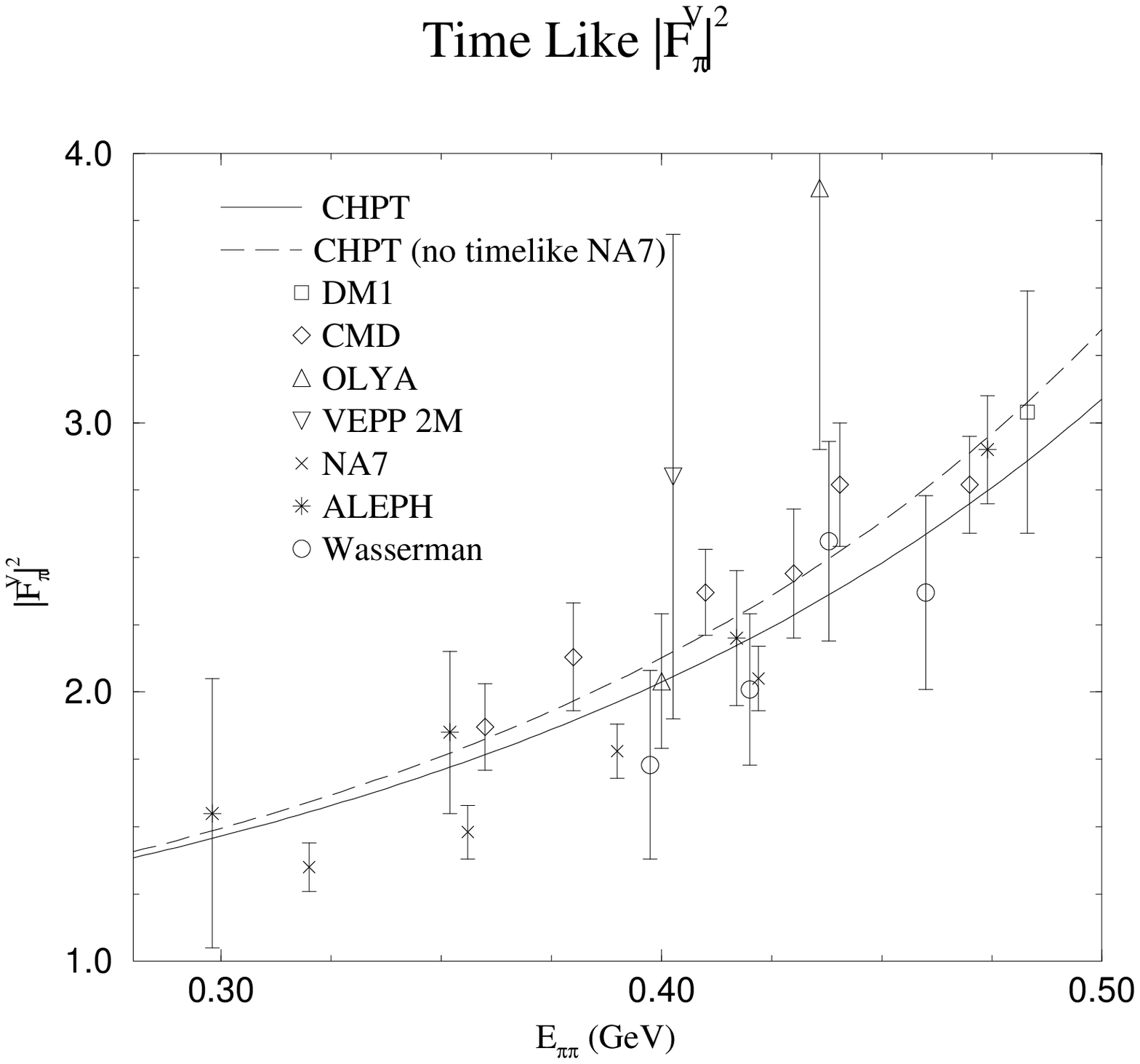}
\caption{\label{timevector} The timelike data and two fits as indicated
  in the caption of Fig. \ref{figFVspace}.} }

\subsubsection{Modified Omn\`es Representation}

In principle we could have tried also for the vector form factor the use of 
a Modified Omn\`es Representation, as we did in the scalar case: we could
have even used that representation to fit the data.
It is clear however that in the vector case the polynomial part is the
largest one, and that exponentiating the small dispersive contribution
would not make a big effect. 
We have actually checked this explicitly and found the effect of the
exponentiation to be rather small up to 700 MeV. 
For this reason we have chosen to estimate the 
effect of higher orders by including an extra term in the polynomial, as
described in the previous subsection.

\subsubsection{Hadronic Contribution to the Muon $(g-2)$ and to
$\alpha(M_Z^2)$}
\label{g-2}

The process $e^+e^-\to\pi^+\pi^-$ is by far the dominant
part of the hadronic cross section at low energies. As was shown in the
previous subsections, we can get a good fit to the data from
Chiral Perturbation Theory up to about 0.5~GeV energies.
We can therefore use our results for an improved estimate of the low energy
hadronic vacuum--polarization contributions to the muon anomalous magnetic
moment and $\alpha(M_Z^2)$. For a recent determination and more
extensive references see \cite{Davieretal}. The relevant formulas are
\ba
\label{integrals}
a_\mu^{had}&=&\frac{\alpha^2(0)}{3\pi^2}\int_{4M_\pi^2}^{\infty}~ds
\frac{R(s)K(s)}{s}\nonumber\\
\Delta\alpha_{had}(M_Z^2)&=&-\frac{\alpha M_Z^2}{3\pi}
~\mbox{Re} \int_{4 M_\pi^2}^{\infty}~ds
\frac{R(s)}{s(s-M_Z^2)-i\epsilon}\,,\nonumber\\
R(s)&=&\frac{3s}{4\pi\alpha^2} \sigma_{\mbox{tot}}(e^+ e^-\to
\mbox{hadrons})
\,.
\ea
The function $K(s)$ is given in \cite{Davieretal}.
As mentioned above, at low energies we have
\be
\sigma(e^+e^-\to\mbox{hadrons})\approx
\sigma(e^+e^-\to\pi^+\pi^-) =
\frac{\pi\alpha^2\left(1-\frac{4 M_\pi^2}{s}\right)^{\frac{3}{2}}}{3s}
\left|F_\pi^V(s)\right|^2\,.
\ee
The total contribution to $a_\mu^{had}$ is dominated by the $\rho$
region with a significant fraction from below 500~MeV.
The contribution to the various quantities as function of the cutoff
$\Lambda^2$ on the integrals in Eq. (\ref{integrals}) is given
in Table \ref{tableamu} for our result for the CHPT form factor
using the fit including all data and the $d_\pi^V s^3$ term.
In brackets we quote the same result but for the fit without the
timelike NA7 data.
The difference is a reasonable estimate for the error involved.
\TABLE{
\begin{tabular}{|c|cc|}
\hline
&& \\
$\; \; \; \; \; \Lambda$ (GeV)$\; \; \; \; \; $ & $\; \; \; \; \;
10^{10}\cdot a_\mu^{had} \; \; \; \; \; $ & $\; \; \; \; \;
10^4\cdot\Delta\alpha(M_Z^2) \; \; \; \; \; $\\ 
&& \\
\hline
&& \\
0.32 & 2.38(2.43) & 0.039(0.040)\\
0.35 & 7.4(7.6)  & 0.13(0.14)\\
0.40 & 20.0(20.6) & 0.42(0.53)\\
0.45 & 35.7(37.2) & 0.86(0.89)\\
0.50 & 53.6(56.3) & 1.45(1.53)\\
&& \\
\hline
\end{tabular}

\caption{\label{tableamu} Contributions of the two--pion production
to $a_\mu^{had}$ and $\Delta\alpha(M_Z^2)$ as a function of the cut--off
$\Lambda$.}
}
This should be compared to the total results from Ref. \cite{Davieretal},
$a_\mu^{had}=(695\pm7.5)\cdot 10^{-10}$ and
$\Delta\alpha^{had}(M_Z^2)=(277.8\pm2.6)\cdot10^{-4}$. 
{}From the present analysis the error of $a_\mu$ coming from the
region below $500$ MeV is 
about $3\cdot10^{-10}$, comparable to the error on the light--by--light
scattering contribution \cite{lightbylight}. Once the $\rho$-mass region
is better explored, more work on both the low energy contribution and the
light--by--light scattering one will be needed.

\section{Conclusions}
\label{conclusions}

In this paper we have calculated the pion scalar and vector form factor
to next--to--next--to--leading order in Chiral Perturbation Theory
and presented simple analytical expressions for all the relevant
quantities. In addition, we have presented the known formulas
for $F_\pi$ and $M_\pi^2$ using the same notation.

We have made a careful comparison of these formulas with the data.
For the scalar form factor this involves a comparison with the form factor
derived using dispersion theory and chiral constraints from
the $\pi\pi$ phase shifts as done in Ref. \cite{DGL}. The CHPT
formula fits well over the entire range of validity. Moreover, we have
shown that by using the ``modified Omn\`es representation'' as proposed in
Ref. \cite{GM} and which aims to resum yet higher orders by exponentiating
part of the unitarity correction, one can improve the chiral
representation, and follow quite closely the exact form factor up to about
700 MeV. 

For the vector form factor we have collected all available data of
reasonable precision and performed first the standard simple fits
to the data sets. Afterwards, we have used the CHPT formula at two loops
together with a phenomenological higher order term to obtain
a new determination of the pion charge radius and $c_V^\pi$:
\ba
\rpiV &=& (0.437\pm0.016)\mbox{ fm}^2 \; \; ,
\nonumber\\
c_V^\pi &=& (3.85\pm0.60)\mbox{ GeV}^{-4} \; \; .
\nonumber
\ea
The error we quote is a combination of theoretical and experimental errors,
it covers the variation of the input parameters over the various fits
and inputs done using the two--loop CHPT formula.

By comparing the Taylor expansions of the measured form factors, and of
their chiral representations, we have been able to better determine some of 
the LEC's that appear in these quantities: two of them are the ${\cal
  O}(p^4)$ constants $\bar{l}_4$ and $\bar{l}_6$, for which we obtained
\be
\bar{l}_4 = 4.4\pm 0.3\quad\mbox{and}\quad\bar{l}_6 = 16.0\pm0.5 \pm 0.7 \; 
\; , 
\ee
where the last error in $\bar{l}_6$ is purely theoretical, and where we
have taken the estimated values of $r_{V1}^r$ and $r_{S2}^r$ into account
in the values given. Notice that $\bar{l}_4$ is practically free from
theoretical uncertainties, as we have shown.
The new value of $\bar{l}_6$ together with the ${\cal O}(p^6)$ results
quoted in \cite{BT} leads to
\be
\bar{l}_5 = 13.0 \pm 0.9 \; \; .
\ee
The other two LEC's that we have determined are ${\cal O}(p^6)$ constants,
that contribute to the quadratic term in the polynomial of the scalar and
vector form factors. We found
\be
r^r_{S3}(M_\rho) \simeq 1.5 \cdot 10^{-4}\; , \; \; \; \; r^r_{V2}(M_\rho)
\simeq 1.6 \cdot 10^{-4} \; \; ,
\ee
with a substantial uncertainty. We find it interesting and encouraging that 
these values are rather close to the estimates we have made on the basis of 
the resonance saturation hypothesis. This result gives support to the idea
that this hypothesis should work even at order $p^6$ of the chiral
expansion. 

\acknowledgments
We thank R. Urech for permission to include parts of \cite{Urech}, and
J. Gasser for enjoyable discussions, and especially for providing us with
numerical values of the exact scalar form factor as given in
Ref. \cite{GM}. 
PT received support from CICYT research project AEN95--0815.
We acknowledge partial support from the EEC--TMR Program, Contract N.
CT98--0169.

\end{document}